\documentclass[preprint,preprintnumbers,amsmath,amssymb]{revtex4}
\usepackage{graphicx}% Include figure files
\usepackage{dcolumn}% Align table columns on decimal point
\usepackage{bm}% bold math

\begin{document}

\title{Enhanced quantum teleportation in non-Markovian environments}

\author{Xiang Hao}

\affiliation{Department of Physics, School of Mathematics and
Physics, Suzhou University of Science and Technology, Suzhou,
Jiangsu 215011, People's Republic of China}

\author{Shiqun Zhu}
\altaffiliation{Corresponding author} \email{szhu@suda.edu.cn}

\affiliation{School of Physical Science and Technology, Suzhou
University, Suzhou, Jiangsu 215006, People's Republic of China}

\begin{abstract}

The protocol of the teleportation of an arbitrary unknown
one-qubit state is investigated when the quantum channel is subject
to the decoherence from the non-Markovian environments with the
memory effects. The quality of the teleportation measured by the
average fidelity and minimal fidelity can be enhanced in the
non-Markovian channel. The memory effects give rise to the revival
of the fidelity which is better than that obtained by the Markovian
channel.

\vspace{1.6cm} Keywords: quantum teleportation, non-Markovian
dynamics, average fidelity


\end{abstract}

\maketitle

\section{Introduction}

The realistic quantum systems\cite{Divin98,Nielsen00} are usually
open to environments which result in the
uncontrollable decoherence and dissipation\cite{Breuer,Zoller}.
Therefore, it is of great importance to study the impacts of
environments on the feasible quantum information processing such as
quantum teleportation \cite{Bennett,Lee00,Bowen,Albeverio,Yeo06}.
Quantum teleportation is the protocol which enables the recreation
of an arbitrary unknown state at a remote place via the local
measurements and necessary classical communications. The key to the
standard teleportation is entangled states acting as the quantum
channel. Recently, some schemes via the noisy channel of the mixed states have been extensively studied in
\cite{Yeo02,Yeo03,Yeo031,Yeo05,Guo06,Zhang07,Zhu05,Oh02,Jung08,Bhaktavatsala}.
These investigations reveal an interesting point that the
entanglement of the channel mainly influences the quality of the
teleportation. The effects of the environments have
also been taken into account. The previous results of the above
works manifest that the type of noises acting on the quantum
channel can determine the efficiency of the teleportation. If the
fidelity is smaller than the critical value of $2/3$, the
entanglement of the channel will be decreased to zero. These
noises are considered in the Markovian limit with the assumption of
an infinitely short correlation time of the environment. However,
the conventionally employed Markovian approximation has faced
more and more challenges because of the rapid advance of
experimental techniques\cite{Breuer}. Recent studies have shown that
non-Markovian quantum processes play an increasingly important role
in many fields of physics \cite{Piilo08,
Piilo09,Yi,Fanchini,Bellomo,Li,Chen}. The apparent feature of the
non-Markovian dynamics is the memory effects from the practical
environment with a certain correlation time. To quantitatively
characterize the memory effect, some measures for the degrees of the
non-Markovianity have been introduced by \cite{Wolf,Piilo091,Xu10}. These
reasons motivate us to investigate how non-Markovian
environments influence a protocol of quantum teleportation.

In this paper, we consider a general mixed two-qubit states as the
noisy channel coupled to two local non-Markovian
reservoirs. The non-Markovian dynamics of the average fidelity of
the teleportaiton is derived. For the long time limit, we use the
minimal fidelity to quantify the success of the teleportation
protocol. Finally, a simple conclusion is given.

\section{The formalism}

In the following discussion, we investigate the quantum teleportation
in the noisy channel which is composed of the two noninteracting
parts. For each part, a qubit $S=A,B$ is locally coupled to a
reservoir $R_{S}=R_{A}, R_{B}$ with the memory effects. The
Hamiltonian of the quantum channel is described by
\begin{equation}
\label{eq:1} H^{c}=\sum_{S=A,B}H_{S}^{0}+H_{R_S}^0+H_{S, R_S}^{I}.
\end{equation}
The inherent Hamiltonian of each qubit $S$ is formed by
$H_{S}^{0}=\omega_0\sigma_{+}^{S}\sigma_{-}^{S}$ where $\omega_0$ is
the transition frequency of the effective two-level system with the
excited state $|e\rangle_{S}$ and ground state $|g\rangle_{S}$.
$\sigma_{\pm}^{S}$ denotes the raising and lowering operators respectively. The
local reservoir is given by the
$H_{R_S}^{0}=\sum_{j}\omega_{j}b_{R_S,j}^{+}b_{R_S,j}$ where the
index $j$ labels the field mode of the reservoir with the
corresponding frequency $\omega_{j}$. $b_{R_S,j}^{+},b_{R_S,j}$ are
the creation and annihilation operators of the mode for the reservoir $R_S$. The
qubit-reservoir interaction Hamiltonian is also obtained by $H_{S,
R_S}^{I}=\sigma_{+}^{S}B_{R_S}+\sigma_{-}^{S}B_{R_S}^{+}$ where
$B_{R_S}=\sum_{j}g_{R_S,j}b_{R_S,j}$ and $g_{R_S,j}$ represents the
coupling between the qubit and the reservoir.

We suppose that the qubits of the channel are initially prepared in
a general mixed states $\rho^{c}(0)$ which can be expressed in the
standard product basis $\{ |1\rangle=|ee\rangle,
|2\rangle=|eg\rangle, |3\rangle=|ge\rangle, |4\rangle=|gg\rangle\}$.
The elements of the density matrix satisfy that $\sum_{n=1}^{4}
\rho_{nn}^{c}(0)=1$ and
$\rho_{mn}^{c}(0)=[\rho_{nm}^{c}(0)]^{\ast}$. If the dynamical map
$\varepsilon_{S}$ for each qubit is known, the evolution of the
states for the quantum channel can be obtained by
\begin{equation}
\label{eq:2}\rho^{c}(t)=\varepsilon_{A}\otimes
\varepsilon_{B}[\rho^{c}(0)].
\end{equation}
In accordance with the result of \cite{Breuer}, the non-Markovian
dynamics $\varepsilon_{S}$ can be written as
\begin{align}
\label{eq:3}\varepsilon_{S}(|e\rangle\langle
e|)&=|G_{S}(t)|^2|e\rangle\langle
e|+[1-|G_{S}(t)|^2]|g\rangle\langle g|& \\ \nonumber
\varepsilon_{S}(|e\rangle\langle g|)&=G_{S}(t)|e\rangle\langle g|&\\
\nonumber \varepsilon_{S}(|g\rangle\langle e|)&=G_{S}^{\ast}(t)|g\rangle\langle e|&\\
\nonumber \varepsilon_{S}(|g\rangle\langle g|)&=|g\rangle\langle
g|,&
\end{align}
where each local reservoir is initially in the vacuum state and the
function $G_{S}(t)=G(t)$ is defined as the solution of the
integrodifferential equation
\begin{equation}
\label{eq:4} \dot{G}(t)=-\int_{0}^{t}f(t-\tau)G(\tau)d\tau.
\end{equation}

The two-point reservoir correlation function $f(t-\tau)$ is closely
related to the spectral density of the reservoir $J(\omega)$,
$f(t-\tau)=\int J(\omega)\exp [i(\omega_0-\omega)(t-\tau)]d\omega$.
For example, we take the detuning case of the Lorentzian spectral
density which describes the vacuum radiation field inside an
imperfect cavity \cite{Breuer},
\begin{equation}
\label{eq:5}J(\omega)=\frac {1}{\pi} \frac {W^2
\lambda}{(\omega_0-\omega-\delta)^2+\lambda^{2}},
\end{equation}
where $\delta=\omega_0-\omega_{c}$ is the detuning between the
center frequency of the cavity $\omega_{c}$ and the resonance
frequency $\omega_0$. $W$ represents the interaction strength
between the qubit and its local reservoir. The parameter $\lambda$
defines the spectral width of the reservoir. It is found out that the effective coupling between the qubit and local reservoir is dependent on the detuning \cite{Breuer}. The large values of $\delta$ represent the weak effective couplings \cite{Breuer}. This also means that the effects of the reservoirs on the protocol are small when the detuning $\delta$ is increased. We can
verify that the correlation function decays exponentially
$f(t-\tau)=W^2\exp[-(\lambda-i\delta)(t-\tau)]$. The correlation
time scale of the reservoir is approximately estimated by
$\tau_{R}\sim \lambda^{-1}$. As another necessary time parameter,
the relaxation time for each qubit is also obtained by $\tau_{S}\sim
\gamma_{0}^{-1}$ where $\gamma_0=2W^2/\lambda$ is regarded as the
decaying rate of the qubit in the Markovian limit of flat spectrum.
Applying the correlation function $f(t-\tau)$, we can obtain the
expression of $G(t)$,
\begin{equation}
\label{eq:6}G(t)=\exp[-\frac 12(\lambda-i\delta)t](\cosh\frac
{dt}2+\frac {\lambda-i\delta}{d}\sinh\frac {dt}2),
\end{equation}
with $d=\sqrt{(\lambda-i\delta)^2-2\gamma_0\lambda}$. If the time
scale $\tau_{R}$ is comparable with the relaxation time scale
$\tau_{S}$, i.e., $\tau_{R}>\tau_{S}/2$, the memory effects of the
reservoir should be taken into account. When $\lambda \ll \gamma_0$ and $\delta=0$, the function is simplified as $G(t)=e^{-\frac {\lambda t}{2}}(\cos \frac{|d|t}{2}+\frac {\lambda}{d}\sin \frac {|d|t}{2})$. It is clearly seen that the revivals of $G(t)$ occur in this case. While $\tau_{R}\ll \tau_{S}$ or $\lambda \gg \gamma_0$, the dynamical map reduces
to the Markovian decoherence where $|G(t)|=e^{-\frac {\gamma_{0}t}{2}}$ is monotonically
decreasing with time.

\section{Teleportation in the environments}

Without the loss of the generality, an arbitrary unknown single
qubit state $|\varphi_{in}\rangle=\cos\frac
{\theta}2|e\rangle+\sin\frac {\theta}2e^{i\phi}|g\rangle$ is
recognized as one input state for the standard teleportation. The
parameter $\theta\in[0,\pi]$ and $\phi\in[0,2\pi]$ represent the
polar and azimuthal angles respectively. The protocol uses a shared
quantum state as a quantum channel to transfer a third quantum state
between two distant parties. For a stationary quantum channel,
quantum teleportation can be performed when the parties $A$ and $B$
can determine which Bell state has the largest overlap with the
quantum channel. When the quantum channel varies with time, the
optimal Bell state will possibly change after the time $T$. In that
time, the quantum channel has evolved to $\rho_c(T)$. If the parties
$A$ and $B$ have complete knowledge of the evolution process, they
will be able to correctly select the optimal Bell state. According
to the work of \cite{Yeo05}, the definite expression of the output
state $\rho_{out}^{(m)}(T)$ can be written as
\begin{equation}
\label{eq:7}
\rho_{out}^{(m)}(T)=\sum_{j=0}^{3}p_{j}^{(m)}(T)\sigma_{j}\rho_{in}\sigma_{j},
\end{equation}
where the local operations $\sigma_{j}(j=1,2,3)$ are the three
components of the Pauli rotation, $\sigma_0=I$ denotes the unity
matrix and $\rho_{in}=|\varphi_{in}\rangle \langle \varphi_{in}|$.
The probabilities $p_{j}^{(m)}= \langle \Psi^{(j\oplus
m)}|\rho^{c}(t) |\Psi^{(j\oplus
m)}\rangle=\mathrm{Tr}[\rho_{Bell}^{(j\oplus m)}\rho^{c}(t)]$
satisfying that $\sum_{j=0}^{3}p_{j}^{(m)}=1$. Although the form of
Eq. (7) is different from the result of \cite{Bowen,Albeverio}, all
of the expressions discover the fact that the teleportation via the
channel of mixed states is equivalent to a generalized depolarizing
channel with the probabilities obtained by the maximally entangled
components of the channel. It is noted that the values of $m$ are
time dependent. Here $j\oplus m (m=0,1,2,3)$ represents summation
modulus $4$ and $\rho_{Bell}^{(j\oplus m)}=|\Psi^{(j\oplus
m)}\rangle \langle \Psi^{(j\oplus m)}|$ describes the density matrix
of the four Bell entangled states $\{ |\Psi^{(0,3)}\rangle=\frac
1{\sqrt 2}(|ee\rangle \pm |gg\rangle) , |\Psi^{(1,2)}\rangle=\frac
1{\sqrt 2}(|eg\rangle \pm |ge\rangle) \}$. The above equation
describes the output state through the channel with the
maximally entangled fraction $|\Psi^{(m)}\rangle $ after the
implementation of Bell measurements and corresponding pauli
operations.

In regard to the environments, the maximally entangled component of
the noisy channel can be varied with time. Therefore, the best
quality of the teleportation can be obtained by the optimal
estimation of the four possible output states. To efficiently
measure the quality of the protocol, the average fidelity between
the input state and output one can be written as $ F^{(m)}=\frac
1{4\pi}\int_{0}^{2\pi}\int_{0}^{\pi}\langle
\varphi_{in}|\rho_{out}^{(m)}|\varphi_{in}\rangle \sin\theta d\theta
d\phi.$ For a given noisy channel, we can optimize the standard
teleportation by choosing certain values of $m$ to maximize the
average fidelity. The achievable maximum of the four average
fidelity can be given by
\begin{equation}
\label{eq:8} F(t)=\frac 13+\frac 23\max_{m}\{ p_0^{(m)}(t)\},
\end{equation}
where the probabilities $p_0^{(m)}(t)=\langle
\Psi^{(m)}|\rho^{c}(t)|\Psi^{(m)}\rangle$ are only connected with
the diagonal and anti-diagonal elements of $\rho^{c}(t)$. The
maximum of $\{ p_0^{(m)}(t),(m=0,1,2,3) \}$ can be obtained as
\begin{equation}
\label{eq:9} \max_{m} \{ p_0^{(m)}(t) \}=\frac 12 \max \{ \mu_1(t),
\mu_2(t) \},
\end{equation}
where
$\mu_1(t)=\rho_{11}^{c}(t)+\rho_{44}^{c}(t)+2|\mathrm{Re}[\rho_{14}^{c}(t)]|$,
$\mu_2(t)=1-\rho_{44}^{c}(t)+2|\mathrm{Re}[\rho_{23}^{c}(t)]|$ and
$\mathrm{Re}[a]$ denotes the real part of $a$. In this case, the
diagonal elements of $\rho^{c}(t)$ are
\begin{align}\label{eq:10}
\rho_{11}^{c}(t)&=\rho_{11}^{c}(0)|G(t)|^4& \\ \nonumber
\rho_{22}^{c}(t)&=\rho_{11}^{c}(0)|G(t)|^2[1-|G(t)|^2]+\rho_{22}^{c}(0)|G(t)|^2&
\\ \nonumber \rho_{33}^{c}(t)&=\rho_{11}^{c}(0)|G(t)|^2[1-|G(t)|^2]+\rho_{33}^{c}(0)|G(t)|^2&
\\ \nonumber
\rho_{44}^{c}(t)&=1-\rho_{11}^{c}(t)-\rho_{22}^{c}(t)-\rho_{33}^{c}(t),&
\end{align}
and the anti-diagonal elements are also given by
\begin{align}
\label{eq:11} \rho_{14}^{c}(t)&=\rho_{14}^{c}(0)G^2(t)& \\ \nonumber
\rho_{23}^{c}(t)&=\rho_{23}^{c}(0)|G(t)|^2.&
\end{align}

For simplicity, we can take the maximally entangled states to be the
initial ones for the channel. When $\rho^{c}(0)=|\Psi^{(0,3)}\rangle
\langle \Psi^{(0,3)}|$, the optimized average fidelity is obtained
by
\begin{equation}\label{eq:12}F(t)=\frac 13[2+|G(t)|^4].
\end{equation}
During the decoherence, the probabilities $|\Psi^{(0,3)}\rangle$ of
the channel are always larger than those of $|\Psi^{(1,2)}\rangle$.
Figures 1-4 show that the dynamics of the optimized average fidelity
is dependent on the scaled time $\gamma_0t$ in the Markovian regime
of $\lambda=5\gamma_0$ and non-Markovian regime of
$\lambda=0.01\gamma_0$. In Fig. 1, the values of $F(t)$
are rapidly decreased to the critical value $2/3$ below which the
quality of the teleportation is worse than that of the classical
communication. With the increasing of the detuning $\delta$, the
decaying rate of the optimized average fidelity is smaller than that
of the resonance case $\delta=0$. In Fig. 2, the revivals of the average fidelity occur after a finite period
time of $F(t)=2/3$ when $\delta=0$. The large detuning of the
reservoir can ensure the preservation of high values of the
optimized average fidelity. This is the reason that the effective interactions between the qubits and reservoirs are weak when the detuning values $\delta$ are large. For $\rho^{c}(0)=|\Psi^{(1,2)}\rangle
\langle \Psi^{(1,2)}|$, the optimized average fidelity is expressed
by
\begin{equation}
\label{eq:17}F(t)=\frac 13+\frac 13\max\{1-|G(t)|^2,2|G(t)|^2 \}.
\end{equation}
From Fig. 3, we can clearly see that the values of the optimized
average fidelity are rapidly decreased to a certain value and then
gradually increased to the critical value $2/3$ in the Markovian
case. Meanwhile, the similar revivals of the average fidelity also
occur in Fig. 4 when the effects of the non-Markovian
environment are considered. When the detuning is large, the
high values of average fidelity can remain during the evolution. In this condition, it is noted that if
the probability $|\Psi^{(1,2)}\rangle$ of the channel is larger than
$|\Psi^{(0,3)}\rangle$, the value of average fidelity $F(t)>2/3$.
This means that the purity of the channel state can affect the
quality of the teleportation.

\section{The minimal fidelity for the protocol}
For the long time, the quantum channel is in the $|gg\rangle $
state. The average fidelity in this case is $2/3$, even though the
$|gg\rangle $ state is nearly the worst quantum channel for
teleportation. From this point of view, we may consider the minimal
fidelity as a better measure to quantify the success of the
teleportation protocol. The minimization can be performed over all
possible values of $\theta$ and $\phi$, i.e.,
$f(t)=\min_{\theta,\phi} \{\langle
\varphi_{in}|\rho_{out}^{(m)}(t)|\varphi_{in}\rangle  \}$. Figure 5
shows the minimal fidelity of the teleportation protocol when the
initial entangled resource is chosen to be $|\Psi^{(0,3)}\rangle$.
The minimal fidelity for $\rho^{c}(0)=|\Psi^{(0,3)}\rangle \langle
\Psi^{(0,3)}|$ can be obtained by,
\begin{equation}
f(t)=\frac 12+\frac {\mathrm{Re}[G^2(t)]}2.
\end{equation}
It is clearly seen that the revival of the fidelity is mainly
dependent on the memory effect from the non-Markovian environment.
For the time limit $t\gg \frac {1}{\lambda}$, the minimal fidelity
for the initial channel of $|\Psi^{(0,3)}\rangle$ is $\frac 12$.
With increasing the detuning value $\delta$, the minimal fidelity
can be greatly improved at the large time scale. This point
demonstrates the memory effects from the non-Markovian environments
can enhance the success of the teleportation protocol. When the
initial channel state is $|\Psi^{(1,2)}\rangle$, the
minimal fidelity is plotted in Figure 6. The analytical expression
of $f(t)$ in this case is given by,
\begin{equation}
f(t)=|G(t)|^2.
\end{equation}
The revival of the fidelity for $\delta=0$ is clearly shown in
Figure 6(a). If the detuning value is increased, the minimal
fidelity can keep the high values for a long time. The large
detuning value helps for the efficiency of the teleportation. In
this case, the values of the minimal fidelity are decreased and
infinitely close to zero with time. This means that the protocol for
the entanglement channel of $|\Psi^{0,3}\rangle$ is much better in
the non-Markovian environment.

\section{Conclusion}

The dynamics of the average fidelity for the standard teleportation
is studied when the quantum channel is subject to the decoherence
from the non-Markovian reservoirs. For the long time limit, the
minimal fidelity is used to quantify the efficiency of the
teleportation protocol. We also investigate the effects of the
non-Markovianity of the channel on the teleportation.
It is also found out that the memory effects can cause the revivals
of the average fidelity and minimal fidelity. Compared with the
Markovian environment, the scheme of the teleportation in the
non-Markovian environment is much better because of the flowing of
information from the environment back to the communication channel.

\vskip 0.5cm

{\large \bf Acknowledgement}

It is a pleasure to thank Yinsheng Ling, Jianxing Fang for their
many fruitful discussions about the topic. The work was supported by
the Natural Science Foundation of China Grant No. 10904104 and No.
11074184.

\newpage

{\large \bf Figure Captions}

\vskip 0.5cm

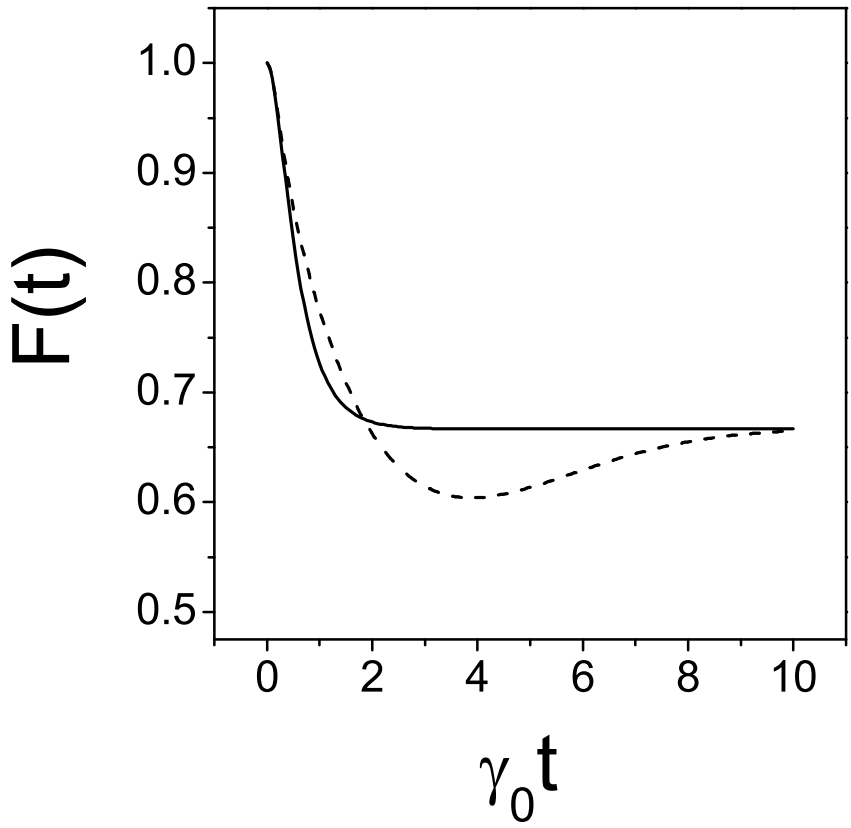
{\bf Figure 1.}

The dynamics of the optimized average fidelity $F(t)$ is plotted as
a function of the scaled time $\gamma_0t$ when the initial channel
state is $\rho^{c}(0)=|\Psi^{(0,3)}\rangle \langle \Psi^{(0,3)}|$.
In the Markovian regime of $\lambda=5\gamma_0$, the solid curve and
dashed curve correspond to the case of $\delta=0$ and
$\delta=5\gamma_0$.

\vskip 0.5cm

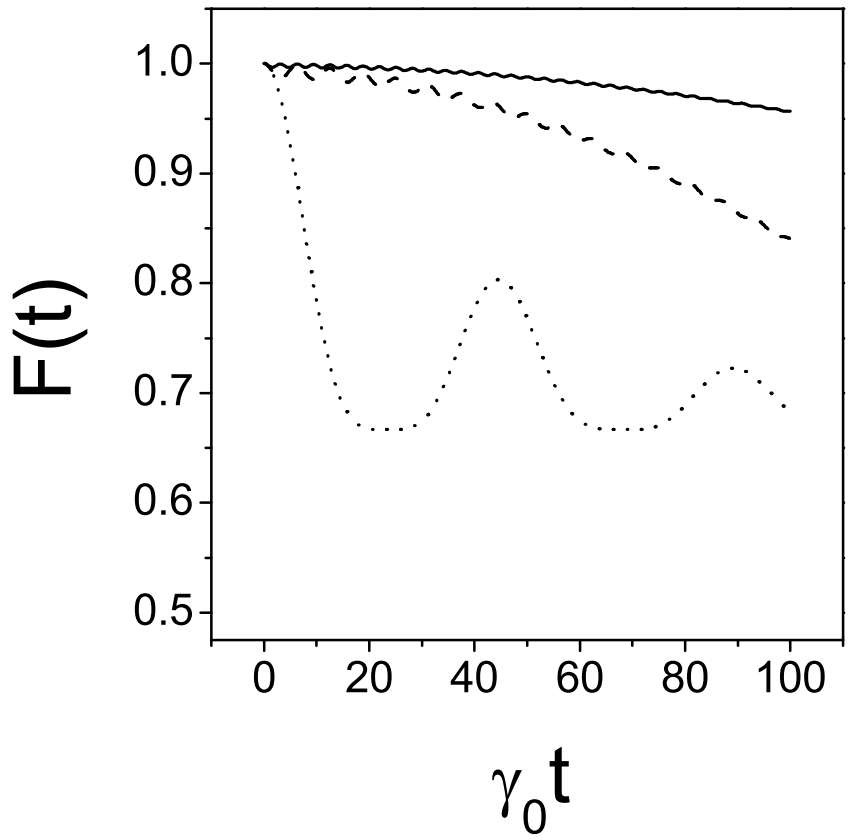
{\bf Figure 2.}

The dynamics of the optimized average fidelity $F(t)$ is plotted as
a function of the scaled time $\gamma_0t$ when the initial channel
state is $\rho^{c}(0)=|\Psi^{(0,3)}\rangle \langle \Psi^{(0,3)}|$.
In the non-Markovian regime of $\lambda=0.01\gamma_0$, the solid
curve, dashed one, dotted curve correspond to the case of
$\delta=2\gamma_0$, $\delta=\gamma_0$ and $\delta=0$.

\vskip 0.5cm

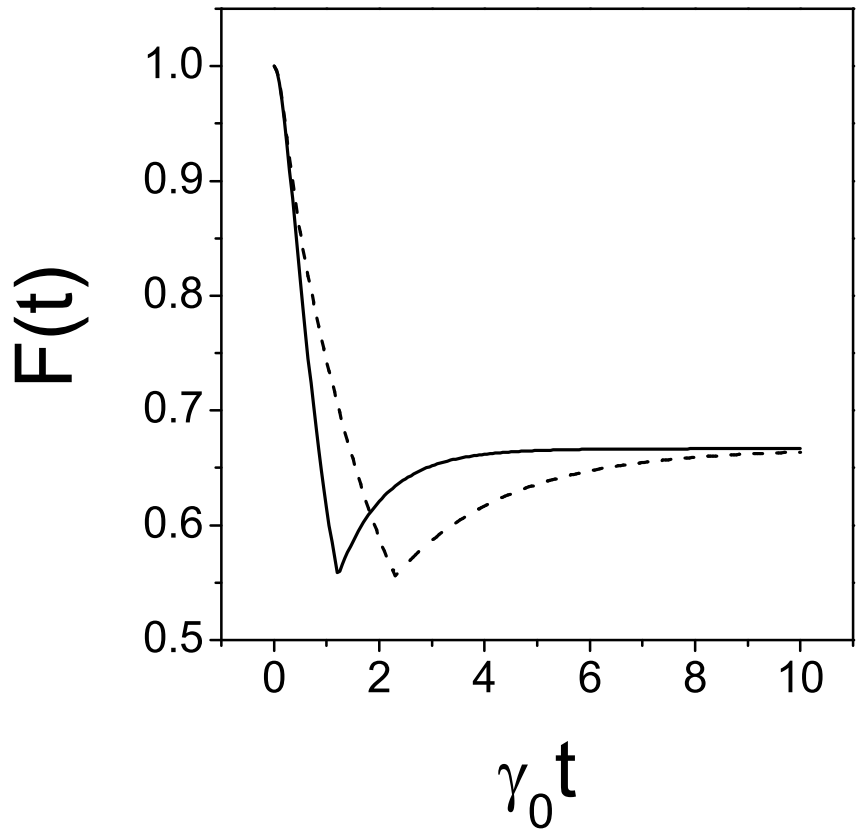
{\bf Figure 3.}

The dynamics of the optimized average fidelity $F(t)$ is plotted as
a function of the scaled time $\gamma_0t$ when the initial channel
state is $\rho^{c}(0)=|\Psi^{(1,2)}\rangle \langle \Psi^{(1,2)}|$.
In the Markovian regime of $\lambda=5\gamma_0$, the solid curve and
dashed curve correspond to the case of $\delta=0$ and
$\delta=5\gamma_0$.

\vskip 0.5cm

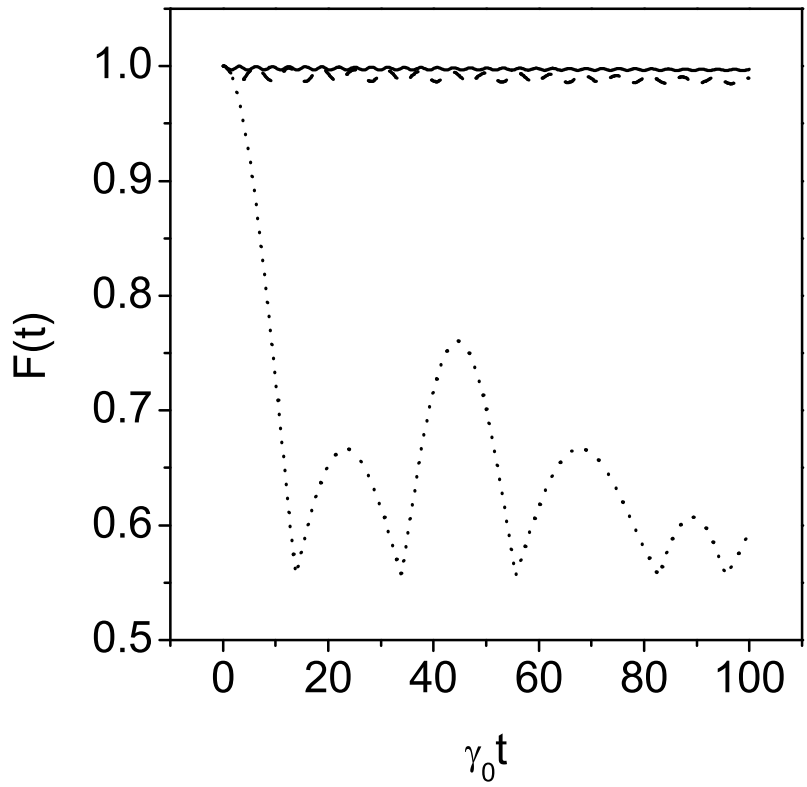
{\bf Figure 4.}

The dynamics of the optimized average fidelity $F(t)$ is plotted as
a function of the scaled time $\gamma_0t$ when the initial channel
state is $\rho^{c}(0)=|\Psi^{(1,2)}\rangle \langle \Psi^{(1,2)}|$.
In the non-Markovian regime of $\lambda=0.01\gamma_0$, the solid
curve, dashed one, dotted curve correspond to the case of
$\delta=2\gamma_0$, $\delta=\gamma_0$ and $\delta=0$.

\vskip 0.5cm

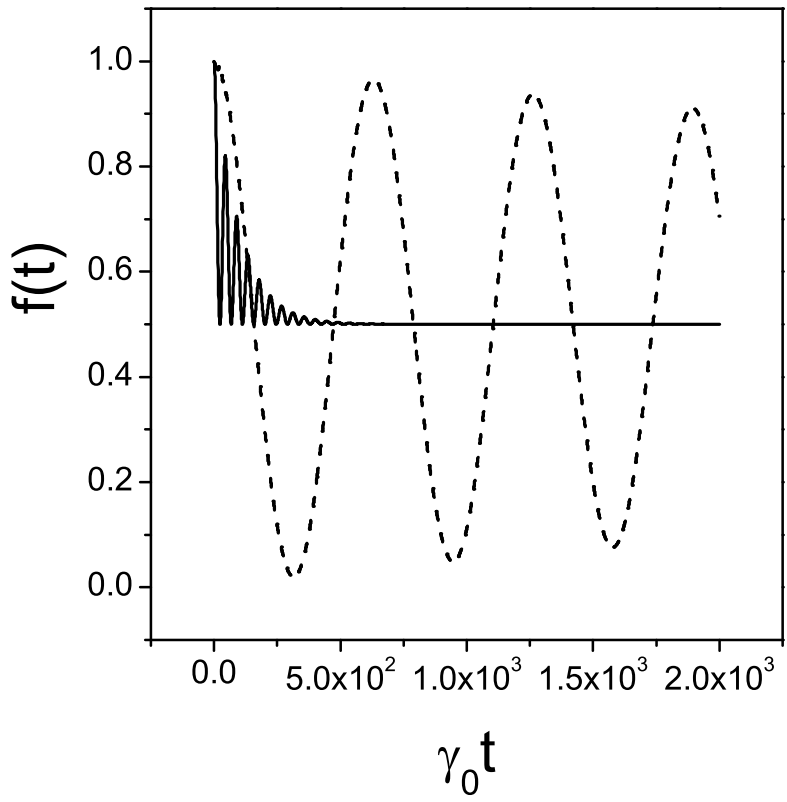
{\bf Figure 5.}

The dynamics of the minimal fidelity $f(t)$ for the initial channel
state of $\rho^{c}(0)=|\Psi^{(0,3)}\rangle \langle \Psi^{(0,3)}|$ is
plotted as a function of the scaled time $\gamma_0t$ when
$\lambda=0.01\gamma_0$. The solid line denots the detuning value of
$\delta=0$. The dashed line represents the case of
$\delta=\gamma_0$.

\vskip 0.5cm

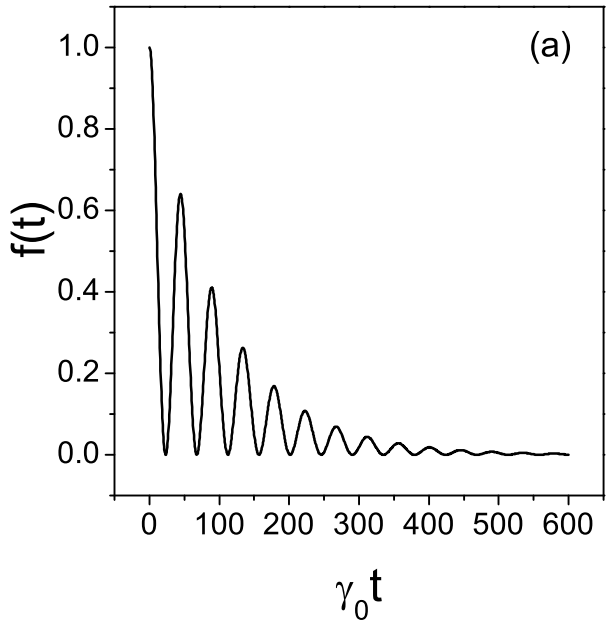
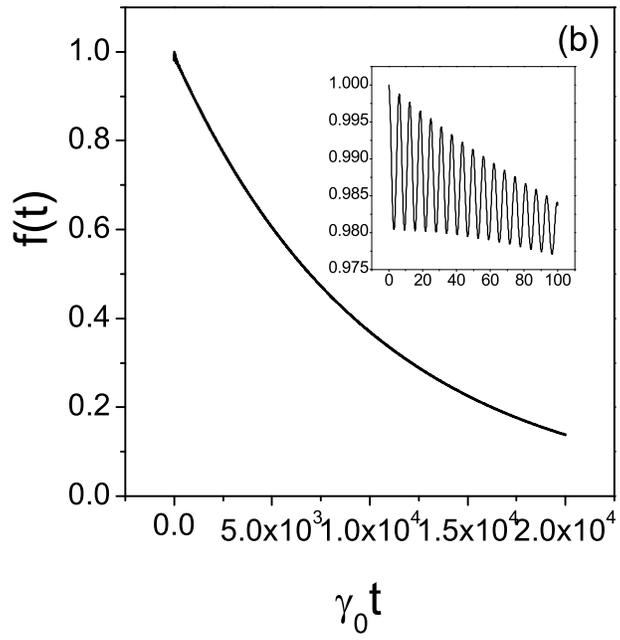
{\bf Figure 6.}

The dynamics of the minimal fidelity $f(t)$ for the initial channel
state of $\rho^{c}(0)=|\Psi^{(1,2)}\rangle \langle \Psi^{(1,2)}|$ is
plotted as a function of the scaled time $\gamma_0t$ when
$\lambda=0.01\gamma_0$. (a) The detuning value $\delta=0$. (b) The
detuing value $\delta=\gamma_0$.

\end{document}